\pdfoutput=1
\documentclass[11pt]{article}
\usepackage[final]{acl}

\usepackage{times}
\usepackage{latexsym}
\usepackage{multirow}
\usepackage[T1]{fontenc}
\usepackage[utf8]{inputenc}
\usepackage{microtype}
\usepackage{inconsolata}
\usepackage{graphicx}

\title{\textit{Retrieve, Annotate, Evaluate, Repeat}: Leveraging Multimodal LLMs for Large-Scale Product Retrieval Evaluation}

\author{
    \textbf{Kasra Hosseini},
    \textbf{Thomas Kober},
    \textbf{Josip Krapac}, 
    \textbf{Roland Vollgraf}, \\
    \textbf{Weiwei Cheng},
    \textbf{Ana Peleteiro Ramallo},
    \\
    \\
    Zalando SE, Berlin, Germany
}

\begin{document}
\maketitle

\begin{abstract}
Evaluating production-level retrieval systems at scale is a crucial yet challenging task due to the limited availability of a large pool of well-trained human annotators. Large Language Models (LLMs) have the potential to address this scaling issue and offer a viable alternative to humans for the bulk of annotation tasks. In this paper, we propose a framework for assessing the product search engines in a large-scale e-commerce setting, leveraging Multimodal LLMs for (i)~generating tailored annotation guidelines for individual queries, and (ii)~conducting the subsequent annotation task. Our method, validated through deployment on a large e-commerce platform, demonstrates comparable quality to human annotations, significantly reduces time and cost, facilitates rapid problem discovery, and provides an effective solution for production-level quality control at scale.
\end{abstract}

\section{Introduction}
Search functionality is a fundamental component of e-commerce platforms, with the objective of finding the most relevant products in a dynamic product database. Customers using search often exhibit a higher intent to find specific products \cite{buying_search_browse}, leading to greater engagement and conversion rates. However, they may struggle to articulate their needs in a search query. Even if they do express their intent clearly, information retrieval~(IR) systems might fail to interpret it correctly, resulting in irrelevant search results \cite{rethinking_e_search}. 

\begin{figure}[ht]
\begin{center}
    \includegraphics[width=1.0\linewidth]{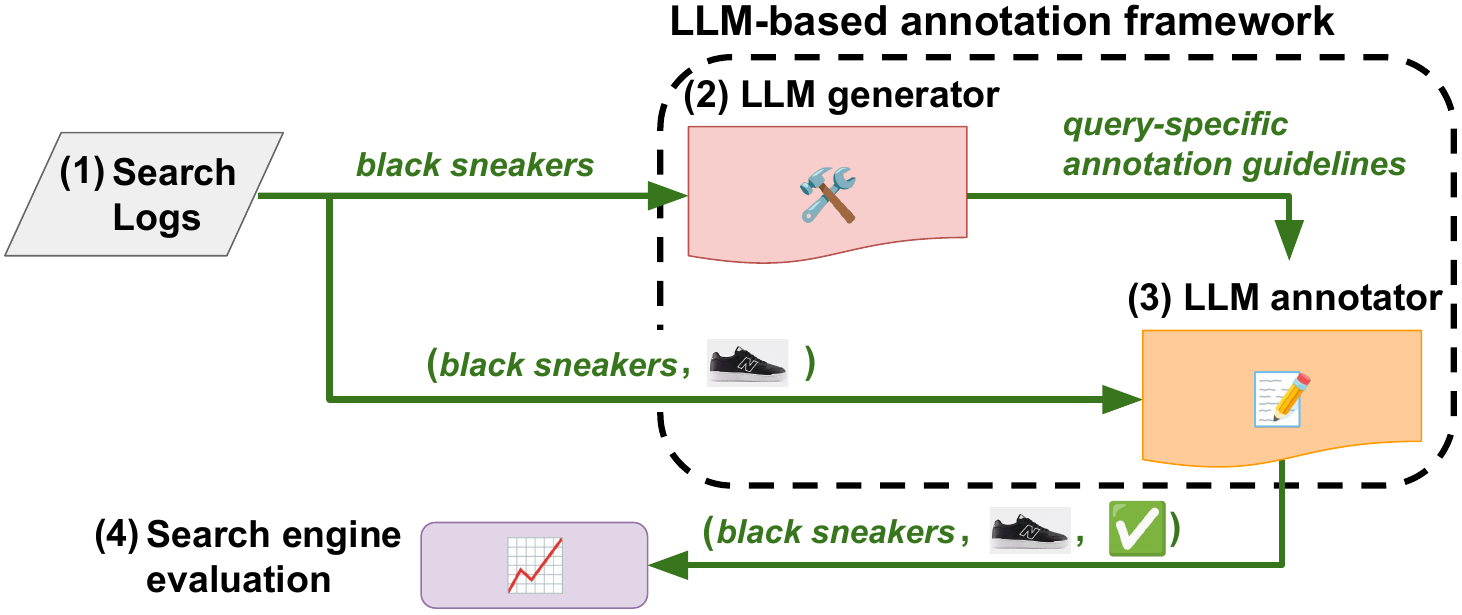}
    \caption{
    Our proposed framework works by extracting a query-product pair from our search query-click logs~(1). The query (e.g. \emph{black sneakers}) is then passed on to the \textbf{LLM generator} (2). The LLM generator creates \emph{specific annotation instructions} for the given query. The query-specific annotation guidelines and the query-product pair (e.g. \emph{black sneakers} and the retrieved product) are provided as input to the \textbf{LLM annotator} (3). Lastly, the annotated query-product pair is forwarded to the search engine evaluation module (4).
    }
    \label{fig:llm_annotation_overview}
\end{center}
\end{figure}

Evaluating product retrieval systems on a large scale in a multilingual setting and for a diverse set of customer queries is an intricate but essential task for maintaining a high-quality user experience and driving business success. A prerequisite for this evaluation is the availability of a large enough pool of query-product relevance labels \cite{voorhees2001philosophy, sigir_2014_workshop}, which indicate whether a retrieved product is semantically relevant to the query. Semantic relevance depends solely on the query and the product, excluding other contextual factors such as personal customer preferences.

Creating annotation guidelines that codify what is semantically relevant is a complex task \cite{spark1975report}. It requires describing the guidelines in a digestible, concise, yet precise manner, as well as curating a set of illustrative examples of varying difficulty. Even with well-defined guidelines and well-trained human annotators, manual annotation is slow and costly.

The advent of crowd-sourcing platforms has increased scalability \cite{blanco2011repeatable,alonso2012using,lease2013crowdsourcing,marcus2015crowdsourced,chen2016introduction}, allowing for a trade-off between speed and cost. However, increasing the number of annotators can lead to inconsistencies, as even the same annotator may provide contradictory annotations for the same query-product pair, let alone multiple annotators. Consistency can be improved by using more annotators per pair (see, e.g., \citet{ferrante2017aware}), but this results in increased cost. In large e-commerce systems, the volume of data that needs to be annotated leads to prohibitively high costs when using crowd-sourcing platforms that rely on human annotators.
 
While the rate of manual relevance judgement varies depending on the task \cite{voorhees2001philosophy,sanderson2010test,chen2022wands,gpt_goes_shopping}, in our use case, we estimate a throughput of 2-3 query-product pairs per minute. As an example, 50,000 queries\footnote{A modest number for the evaluation of large IR systems.} and 20 products per query results in one million query-product pairs, which takes 5,500-8,500 hours of human labour, assuming one annotation per pair. Moreover, evaluation is not a one-off practice; ideally, companies continuously assess their search engines to ensure effectiveness over time.

The sheer volume of required annotations in multiple languages, along with the need for continuous evaluation, makes human-generated relevance judgements the primary bottleneck in creating product retrieval evaluation datasets. To overcome these challenges, there has been growing interest in leveraging LLMs \cite{perspectives_LLM_relevance,thomas2023large,gpt_goes_shopping,rahmani2024synthetic,upadhyay2024llms,bergum2024improving}.

In this study, we propose a framework that leverages the capabilities of Multimodal Large Language Models (MLLMs) for assessing the relevance of query-product pairs (Fig.~\ref{fig:llm_annotation_overview}). Our method combines the strengths of LLMs and MLLMs in understanding natural language queries across various languages and processing both textual and visual features of products. Unlike traditional per-task annotation guidelines, such as those discussed by \citet{gpt_goes_shopping}, we employ LLMs to generate annotation guidelines \emph{specific to each query}. Additionally, our pipeline's modular design allows for caching and parallel processing, which is crucial for scaling up to larger systems. This framework has enabled daily evaluations of our product retrieval systems.
It has also facilitated the comparison of different search models, increasing our confidence in offline evaluations and complementing our online evaluation techniques, such as A/B testing and other controlled online experiments \cite{kohavi2009controlled}. 
Moreover, we have used the relevance assessments' outputs to train, evaluate and analyse other components of our search and ranking systems.\\We furthermore show that while human-human and human-LLM agreement scores are on par with each other, we find that humans and LLMs tend to make very different types of annotation errors. Our findings suggest that LLMs are very effective for the bulk work of annotations whereas human expertise is better leveraged for more complex cases. 

In summary, our contributions are as follows:

\begin{itemize}
    \item We introduce a multimodal LLM-based evaluation framework for large-scale product retrieval systems. We propose \textit{query-level} annotation guidelines and utilise multimodal inputs (text and images) for relevance assessment.
    \item We evaluate the performance of our framework against human annotations on real-world production search queries in a multilingual setting and analyse the different types of errors that humans and LLMs tend to make.
    \item We demonstrate the cost-effectiveness and efficiency of our approach for conducting large-scale evaluations. We also compare the performance of different types of LLMs \cite{radford2019language,brown2020language,achiam2023gpt} for relevance assessment.
\end{itemize}

\section{Multimodal LLM-based relevance assessment}
\label{sec:pipeline}

\begin{figure*}[ht]
\begin{center}
    \includegraphics[width=1.0\linewidth]{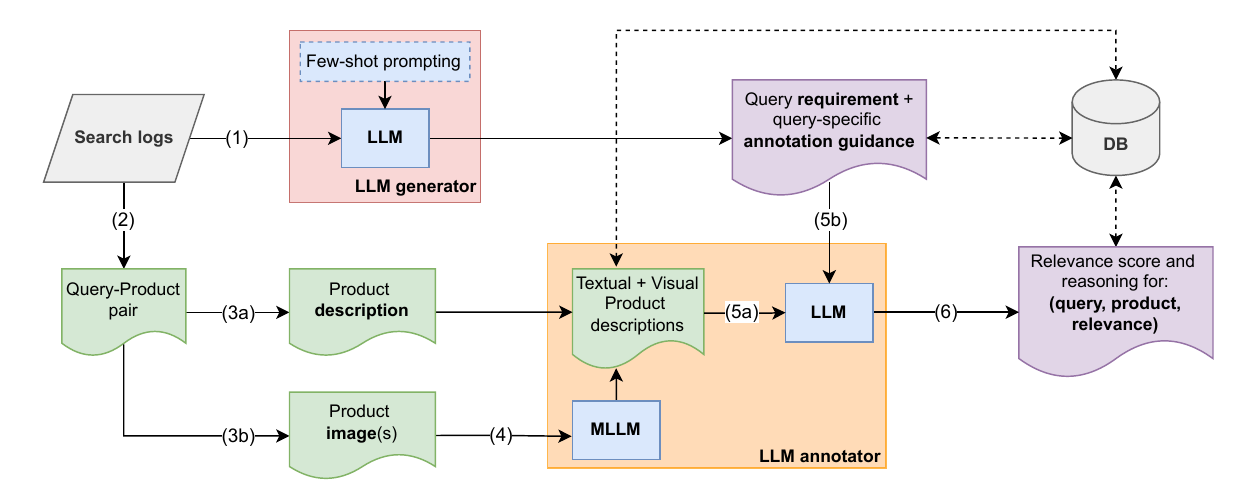}
    \caption{
    Our proposed Multimodal LLM-powered framework enables offline evaluation of large-scale product retrieval systems and presents significant time and cost reductions compared to existing evaluation techniques. Refer to Fig.~\ref{fig:llm_annotation_overview} for an overview of the main steps in the framework, and consult the text for further details. 
    The orange rectangle indicates where a ``one-step'' Multimodal LLM (MLLM) could be utilised, instead of employing one MLLM to create a textual description for image inputs (Step~4) followed by an LLM (Step~5a). In the one-step MLLM, both textual descriptions and the product image are directly fed into the LLM annotator, along with query requirements and query-specific annotation guidelines. The depiction of the pipeline is simplified for readability.
    }
    \label{fig:pipeline}
\end{center}
\end{figure*}

The setup of our method is depicted in Fig.~\ref{fig:pipeline}. It is designed to leverage the capabilities of (M)LLMs for efficient evaluation of large-scale product retrieval systems, and it consists of six main steps: \\
(1) For a given query and its context (e.g., selected gender and market), an LLM generates a query requirement list and a \textit{query-specific} annotation guideline. The query requirement list captures the relevant pieces of information in the user's query and their level of importance. For example, for the query \textit{Nike red shoes}, the query requirement list includes the brand (\textit{Nike}), colour (\textit{red}) and product category (\textit{shoe}). The query-specific annotation guideline is generated by the LLM based on the query and its requirement list. It outlines criteria for each predefined label (see Appendix~\ref{app:evaluation_example_steps} for a detailed example). In our experiments, we defined three relevance labels for a query-product pair: ``irrelevant'', ``acceptable substitute'' and ``highly relevant''.\footnote{See Appendix~\ref{app:annotation_guidelines} for details.} \\ \\
(2) The query and its context are sent to the search engine, which retrieves a set of products. For simplicity, we illustrate this process using a single query-product pair. However, in practice, we work with multiple query-product pairs and may utilise two or more retrieval systems, particularly when comparing their performance. \\
(3a,b) For each retrieved product, we have access to its textual description and its associated image. \\
(4) Using MLLMs and the product image, a visual description in textual form is generated. \\
(5a,b) The combined textual and visual product descriptions are sent to an LLM together with the outputs of Step~1 (i.e., query requirement list and query-specific annotation guideline). \\
(6) The LLM assigns a relevance score to the query-product pair using a set of predefined labels. In its simplest form, the output is a database with one row for each \texttt{(query, product, relevance score)}.\\
In  Steps~1 and 6, we utilise chain-of-thought~(CoT) prompting \cite{wei2022chain,nye2021show} to enhance the quality of (M)LLM outputs and for debugging. An example of the reasoning steps is shown in Appendix~\ref{app:evaluation_example_steps}.\\
As illustrated with dashed lines in Fig.~\ref{fig:pipeline}, all outputs and intermediate steps are stored in a database. This caching serves two key functions in our pipeline. Firstly, it facilitates efficient retrieval and reuse. When evaluating a new search engine configuration (or a variation of existing ones), the database is queried to retrieve relevant pieces of information, including the query requirement list, query-specific annotation guidance, textual and visual product descriptions, and relevance scores. We only compute the missing pieces of information. Secondly, it ensures consistent evaluation across different search engines, as intermediate steps (such as query-specific annotation guidelines) are computed only once and then used to evaluate various search engines.

\section{Experiments and Results}
\label{sec:experiments}

\subsection{Dataset}
\label{sec:exp_dataset}

\textbf{Data collection.} As a starting point for our data collection, we used one year’s worth of production search query traffic\footnote{Our data collection process complies with the regulations defined in the GDPR and other existing regulatory frameworks around data privacy and safety in the European Union.}. We then performed stratified sampling along the following dimensions: a) search engine, b) activated gender filter on the website, c) query frequency, and d) query length in tokens.

\begin{table}[htbp]
\centering
\caption{Dataset statistics.}
\resizebox{\columnwidth}{!}{
\begin{tabular}{ l | c | c | c | c }
    \textbf{Language} & \textbf{Unique} & \textbf{Unique} & \textbf{Avg. tokens} & \textbf{Unique} \\
    & \textbf{pairs} & \textbf{queries} & \textbf{per query} & \textbf{products} \\
    
    \hline
    German & 10,000 & 500 & 3.68 & 8,076 \\
    English & 10,000 & 500 & 3.99 & 8,652
\end{tabular}}
\label{tab:dataset_statistics}
\end{table}

\begin{figure*}[ht]
\begin{center}
    \captionsetup{type=table}
    \caption{
    Agreements between (M)LLM and the human annotator groups (i.e., A1 and A2). We compare agreements based on i) matching either A1 \emph{or} A2 and ii) inter annotator agreement between human annotators (A1 vs. A2) and between LLMs and the human majority vote.
    In the A1 \emph{or} A2 column, we use the same human majority vote to measure the agreements for human annotators.
    Results are reported separately for English and German. For human annotations, we report the total time and cost. We use GPT-4o in all steps of our LLM annotation pipeline (Fig.~\ref{fig:pipeline}).
    Refer to Table~\ref{tab:llm_vs_human_gpt4o_with_without_more_columns} for a more detailed comparison between human annotator groups (A1, A2, and tiebreaker) and different versions of our LLM-powered framework.
    }
    \includegraphics[width=0.8\linewidth]{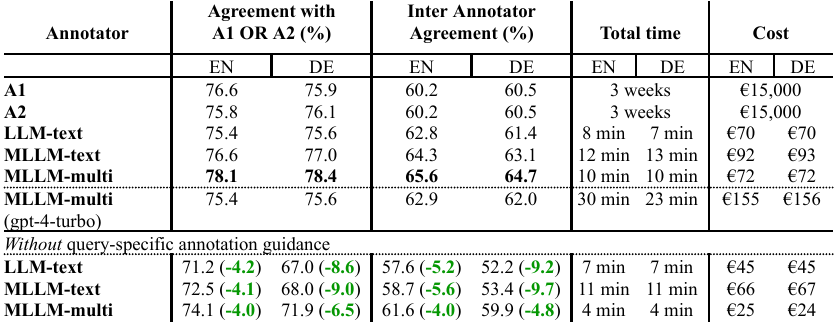}
    \label{tab:llm_vs_human_gpt4o_with_without}
\end{center}
\end{figure*}

After sampling and manual curation\footnote{We manually checked all sampled queries to ensure we cover a diverse and broad spectrum. For example, we would manually replace \emph{yet another} sneakers-related query with a boots-related query of the same length.}, we obtained 500 queries in German and English each. For each query, we then used the existing search engine to retrieve 20 products, selecting 15 products from the top of the retrieved results and randomly sampling the remaining 5 from rank 500 onwards. This resulted in 10,000 unique query-product pairs for German and English each. Table~\ref{tab:dataset_statistics} outlines the statistics of the collected dataset.

\textbf{Data annotation.}~For data annotation, we contracted an external agency to facilitate crowd-sourcing annotations based on the guidelines we provided.\footnote{See Appendix~\ref{app:annotation_guidelines} for an overview of our annotation guidelines.} Our requirements specified that only native speakers (German and English, see Table~\ref{tab:dataset_statistics}) annotate the data. Each query-product pair was to be annotated by two human annotators, with an additional third annotation as a tiebreaker in cases where the two original annotators disagreed. The data annotation process was done in three phases: (i)~a pilot phase to onboard annotators and resolve outstanding loopholes and questions in the annotation guidelines; (ii)~the main annotation phase; (iii)~the tiebreaker phase.\\In total, the data annotation process – from the handover of the initial version of the annotation guidelines by us to the final delivery of annotated data by the external agency – took approximately 8 weeks\footnote{This excludes the effort needed to write the annotation guidelines and to sample the query-product candidate pairs.}, with roughly 3 weeks needed for completing the main annotation and tiebreaker phases. The total cost of data annotation was €30,000.

\subsection{LLM versus Human Annotators}
\label{sec:llm_vs_human_annotators}

Using our proposed framework, we assessed the relevance quality of the 20,000 unique query-product pairs. Table~\ref{tab:llm_vs_human_gpt4o_with_without} summaries the results of our experiments using few-shot prompting, where we incorporated examples into the system prompts of the (M)LLMs. Here, we provided the LLMs with example customer queries, their corresponding requirement lists, and quality labels, but not complete product descriptions or images.\\
Initially, we randomly sampled 100 examples from the English dataset and examined the relevance labels assigned by both LLMs and human annotators. We used the results of this step to adjust the few-shot examples in the system prompt. 

We compare the performance of different versions of our pipeline with human annotations. In Table~\ref{tab:llm_vs_human_gpt4o_with_without}, these versions are labelled as ``LLM-text'', ``MLLM-text'', and ``MLLM-multi''.\\``LLM-text'' is the simplest version where only product descriptions in textual form are used, without incorporating product images. In ``MLLM-text', we employ a vision model to generate textual descriptions of product images (Step~4 in the pipeline, see Fig.~\ref{fig:pipeline}). The generated textual description of the product image is then concatenated with the product description itself (Step~5a). ``MLLM-multi'' utilises the same textual input as ``LLM-text'', while also incorporating the product image as an additional input.\footnote{See the orange rectangle in Fig.~\ref{fig:pipeline} where both the textual product description and its image are fed into an MLLM.} Comparing ``MLLM-multi'' and ``LLM-text'' highlights the impact of multimodal inputs on our task.\\In all cases, the (M)LLM uses product information (in different modalities, depending on the version), query requirements, and query-specific annotation guidance to assign relevance labels.\\Overall, Table~\ref{tab:llm_vs_human_gpt4o_with_without} 
shows that the agreement between human annotators and LLMs is on par with that between human annotators, supporting the scalability of LLM annotation for production-level traffic.

Table~\ref{tab:llm_vs_human_gpt4o_with_without} also shows the results of an ablation study that removed the query-specific annotation guideline (Step~1 in Fig.~\ref{fig:pipeline}). The inclusion of this guideline improved agreements by approximately 4-10\%. More importantly, this component in our framework is essential for enhancing the interpretability and debugging of LLM-based decisions. However, as expected, incorporating query-specific annotation guidelines and chain-of-thought reasoning increased the evaluation costs.\\
We also tested the impact of different (M)LLM architectures in our pipeline.\footnote{Refer to Table~\ref{tab:llm_vs_human_gpt4o_with_without_more_columns} and~\ref{tab:gpt35_4o_for_vision} in Appendix~\ref{app:other-experiments-llm-type} for the results of similar experiments conducted with different LLM types.} The results shown in Table~\ref{tab:llm_vs_human_gpt4o_with_without} are based on ``GPT-4o'' \cite{openai_gpt4o} except for one row, labelled ``MLLM-multi (gpt-4-turbo)''. In the case of GPT-4 Turbo, the agreement with human annotators consistently fell below that of GPT-4o, while its costs and evaluation times exceeded those of all other architectures.\\
In Table~\ref{tab:gpt35_4o_for_vision}, we repeated the experiments using GPT-3.5 Turbo. As expected, the results were significantly worse compared to GPT-4o or GPT-4 Turbo. However, the cost and time required for GPT-3.5 Turbo were lower than for the other architectures.

\section{Discussion}
\label{sec:discussion}

\textbf{Agreement between LLM and human annotators.} The human-LLM agreements between ``MLLM-multi'' and the human majority vote – 65.6\% for EN and 64.7\% for DE in Table~\ref{tab:llm_vs_human_gpt4o_with_without} – are in line with the human inter-annotator agreement, which is 60.2\% for EN and 60.5\% for DE. 

To better identify discrepancies between LLM and human annotations, we focused our analysis on hard disagreements between the two. We consider a hard disagreement to be when, for example, the LLM considers a product to be ``highly relevant'' for the given query, whereas the human majority judgement would be ``irrelevant'', and vice versa. In total, we found that approximately 15\% of annotations\footnote{2,971 out of 20,000 query-product pairs ($\approx$15\%) had hard disagreements between the LLM and human majority vote.} in our dataset were hard disagreements. For manual analysis, we sampled 20\% of the hard disagreements and found that in 50\% of the cases, the human annotation was wrong, in 31\% the LLM was wrong, and in 19\% of cases, both the LLM and the humans provided a wrong annotation.

\begin{figure}[ht]
\begin{center}
    \includegraphics[width=1.0\linewidth]{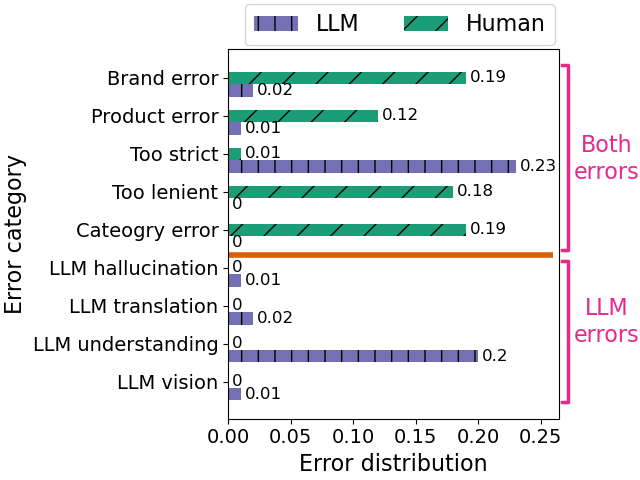}
    \caption{
    Distribution of errors between LLMs and humans on hard disagreements (50\% were due to human errors, 31\% LLM errors and in 19\% both made an error). The upper part (``Both errors'') focuses on errors that either the LLM or humans could make. It highlights that LLMs and humans make very different types of errors. In addition, the lower part (``LLM errors'') shows the distribution of errors that only an LLM would make. Predominantly these are misunderstandings of a part of the search query.
    }
    \label{fig:llm_vs_human_error_analysis}
\end{center}
\end{figure}

We further categorised the hard disagreements into 9 error classes\footnote{See Appendix~\ref{app:llm_vs_human_error_types} for the definition of these classes and for example errors we have observed.}, and found LLMs and humans tend to make very different kinds of errors. For example, as shown in Fig.~\ref{fig:llm_vs_human_error_analysis}, the main errors made by the LLM are either being \emph{too strict} in their judgement (e.g., considering a product as ``irrelevant'', where ``acceptable substitute'' would have been more appropriate), or misunderstanding a part of the query (e.g., interpreting \emph{On Vacation} in its literal sense rather than the fashion brand). On the other hand, humans would oftentimes be \emph{too lenient} when LLMs were \emph{too strict} (e.g., considering a product as ``highly relevant'' when ``acceptable substitute'' would have been more appropriate). Furthermore, human annotations frequently exhibited brand errors (e.g., considering a pair of \emph{Lee} jeans as ``highly relevant'' for a query requesting \emph{Levi's jeans}), product errors (e.g., considering an \emph{Adidas Samba} sneaker as ``highly relevant'' for a query requesting an \emph{Adidas Stan Smith} sneaker), or category errors (e.g., considering a pair of \emph{Nike shirts} as ``highly relevant'' for a query requesting \emph{Nike shoes}), which we barely ever observed for LLMs. We hypothesise that the latter three kinds of human errors are primarily due to annotation fatigue as specifically these cases have been prominently and unambiguously featured in the annotation guidelines. 

These findings suggest that LLMs might be a more reliable source for the bulk of annotations, freeing human labour to focus on trickier cases.\footnote{This typically includes new trending terms or styles that LLMs have not observed yet, but also ambiguous queries such as \emph{old money} or \emph{dark academia}, that specify a fashion style.} In the human-machine collaboration spectrum introduced by \citet{perspectives_LLM_relevance}, our approach can be classified as a ``Human Verification'' (or human-in-the-loop) approach.

\textbf{Subjective nature of relevance judgements.} We found that human disagreement was dominated by two main factors, (i) human errors due to annotation fatigue as described above\footnote{In fact we found that human-human hard disagreements, also making up approx. 15\% of the annotated data, were also primarily due to brand errors, product errors and category errors --- the same types of errors that we also found when comparing LLM and human hard disagreements.}, and (ii) the inherent subjective nature of the task. For the latter, we attribute the source of disagreement to either the ambiguity in the annotation guidelines (even comprehensive guidelines cannot cover all possible cases), or to the subjective judgement of the annotator.\footnote{See Appendix~\ref{app:subjectivity} for examples.} Ideally, the annotation guidelines should make the task as objective as possible; however, in practice, there is always a level of subjectivity.

\textbf{Annotation time and cost.} (M)LLMs are approximately 100 to 1,000 times cheaper than human annotators, and the time required to complete all 20,000 annotations using (M)LLMs is significantly smaller (around 20 minutes for (M)LLMs compared to about 3 weeks for human annotators). Note that several human annotators worked in each group (i.e., ``A1'' and ``A2'' in Table~\ref{tab:llm_vs_human_gpt4o_with_without}), and the total time reported in Table~\ref{tab:llm_vs_human_gpt4o_with_without} is for annotating all query-product pairs. This excludes the time spent on scoping and onboarding human annotators. For (M)LLMs, the reported time excludes the pipeline development time and only includes the actual annotation time.

We anticipate that both cost and time will decrease even further as LLMs and their APIs become more efficient. Moreover, new approaches, such as batch processing, can further reduce costs (e.g., OpenAI's new batch processing is half the price of non-batch queries\footnote{\url{https://platform.openai.com/docs/guides/batch} (accessed on 2024-07-17)}). Indeed, in production, we use batch processing to assess query-product pairs across markets on a nightly basis.

\textbf{Relevance assessment in production.}
High relevance is a necessary, but not a sufficient condition, for high customer engagement, as it is also determined by other factors, e.g. personal preferences, product availability, and price expectations. In this paper, we focus on semantic relevance, but in production we rank the retrieved documents based on various features to take into account both relevance to the query and customers' personal preferences.

Currently, we use the LLM-powered evaluation framework presented in this paper in production to continuously perform relevance assessments at scale.
We typically focus on monitoring the performance of high-volume queries with our framework. Additionally, we evaluate the retrieval performance for low-performing queries. We identify such queries based on signals indicating low relevance in top ranked results, such as low engagement with the result set and high friction in customer experience (e.g., a high reformulation rate\footnote{The reformulation rate is the percentage of queries that are modified and resubmitted within the same search session, indicating an initial failure to satisfy the user's intent and a subsequent attempt to refine the search.}) or high exit rate. This approach enables us to significantly reduce costs and to enhance customer experience faster by prioritising the queries that need the most attention and optimising our resources accordingly.

\section{Conclusion}
Our novel evaluation method leveraging Multimodal LLMs demonstrates a highly efficient approach to assessing large-scale IR systems in product retrieval. We introduce query-level annotation guidelines for calibration and utilise the multimodal capabilities of foundation models to assess the relevance of retrieved products for a query. Our LLM-powered framework, combined with caching and parallel processing, leads to significant reductions in both time and cost. The method's scalability, ability to handle multilingual queries and products, and support for continuous offline evaluations are crucial for large IR systems operating in diverse markets. Experimental results, validated against 20,000 human annotations, confirm the effectiveness and efficiency of our approach. A detailed analysis of human and (M)LLM annotations indicates that (M)LLMs are a more reliable source for relevance assessment in large-scale IR systems. We are currently leveraging this framework in production to continuously perform relevance assessments at scale and maintain a high-quality user experience. Additionally, we utilise its outputs to train, evaluate, and analyse other components of our search and ranking systems.

\section{Ethics Statement}
Our data collection process strictly adheres to the General Data Protection Regulation (GDPR) and other relevant data privacy and safety laws within the European Union. We ensure that all data utilised, including human evaluation data, is anonymised to safeguard against the disclosure of any personally identifiable information.

We do not suggest replacing human annotators with large language models (LLMs). Instead, we focus on leveraging the strengths of both. Our analysis indicates that human annotators may make errors due to annotation fatigue or lack of domain knowledge—errors not observed with LLMs. Therefore, we recommend using LLMs for bulk annotation work while reserving human expertise for more complex cases. 

We are committed to advancing responsible and unbiased AI technologies and welcome any inquiries regarding the ethical aspects of our work.

\bibliography{paper}

\clearpage

\appendix

\section{Multimodal LLM-powered relevance assessment: evaluation steps for an example query}
\label{app:evaluation_example_steps}

\begin{figure*}[ht]
\begin{center}
    \includegraphics[width=1.0\linewidth]{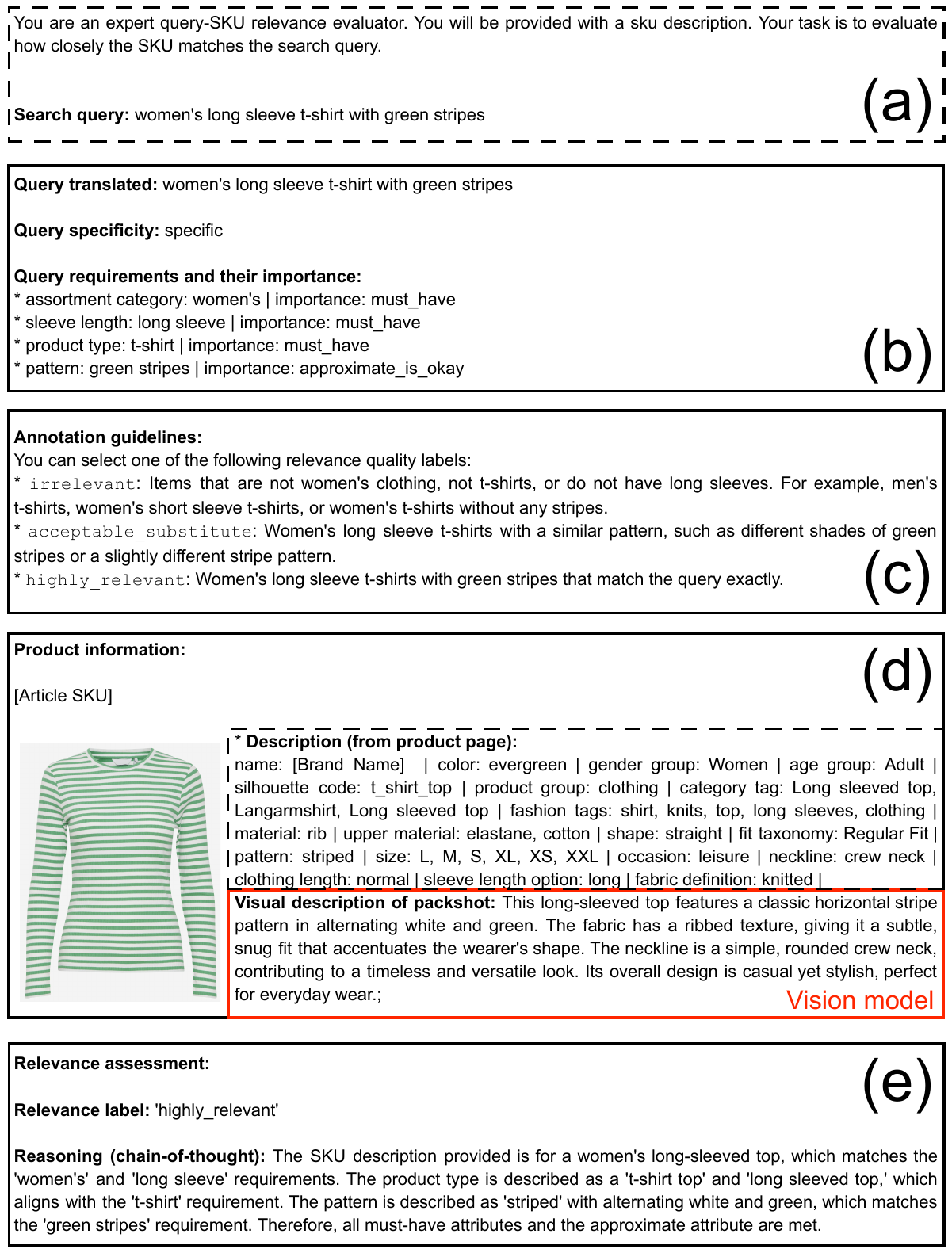}
    \caption{Evaluation steps for an example query \textit{women's long sleeve t-shirt with green stripes}. The entire content displayed in this figure is generated by Multimodal LLMs, except for panel (a), the packshot in panel (d), and the black dashed rectangle also in panel (d). However, within the attributes shown in panel (d), the ``visual description of packshot'', highlighted by a red rectangle, is also generated by a vision model (specifically, GPT-4o was used in this instance). Please refer to the text for further details. (In this example, we have removed the brand name from the product description and the tag on the packshot.).}
    \label{fig:evaluation_steps_for_example}
\end{center}
\end{figure*}

\begin{figure*}[ht]
\begin{center}
    \includegraphics[width=1.0\linewidth]{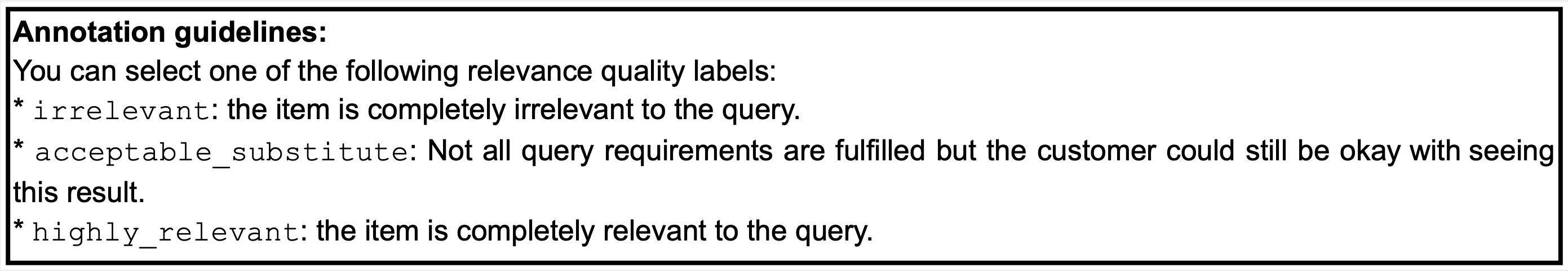}
    \caption{A \textit{generic} annotation guideline for the task of query-product relevance assessment. Compare this to the \textit{query-specific} annotation guidelines in panel (c) of Fig.~\ref{fig:evaluation_steps_for_example}.}
    \label{fig:evaluation_steps_for_example_generic}
\end{center}
\end{figure*}

Fig.~\ref{fig:evaluation_steps_for_example} illustrates the various steps of our evaluation framework using the example query \textit{women's long sleeve t-shirt with green stripes}.\\
Given this query, the LLM infers its requirements and their importance (Step~1 of our framework, refer to Fig.~\ref{fig:pipeline}). The outputs of this step for the example query are detailed in the paragraph ``Query requirements and their importance'' in Fig.~\ref{fig:evaluation_steps_for_example}b. As shown in Fig.~\ref{fig:evaluation_steps_for_example}b, the LLM has inferred four query requirements: ``assortment category'', ``sleeve length'', ``product type'' and ``pattern''. An importance level is also assigned to each requirement (in this case, the first three requirements are ``must\_have'', and the last one is ``approximate\_is\_okay''). Additionally, the LLM provides a reason for each requirement and its importance (not shown here). The LLM also translates the query into English and assigns a ``specificity'' level, as shown in Fig.~\ref{fig:evaluation_steps_for_example}b. \\
In panel (c), the translated query, its specificity, its requirements and their importance are used to create \textit{query-specific annotation guidelines}. The three quality labels (i.e., ``irrelevant'', ``acceptable\_substitute'' and ``highly\_relevant'') are pre-defined. However, the guidance for each label is generated by the LLM. The LLM provides clear and detailed descriptions for each relevance label, tailored to the given query. In the ablation study of Table~\ref{tab:llm_vs_human_gpt4o_with_without}, we assessed the impact of query-specific annotation guidelines on our method's performance. To do this, we replaced the query-specific guidelines with a generic one, as shown in Fig.~\ref{fig:evaluation_steps_for_example_generic}.\\
In Fig.~\ref{fig:evaluation_steps_for_example}d, an example product, its attributes, and its image are shown. These attributes are read from an existing database and are not generated by the LLM, except for the ``visual description of packshot'', highlighted by a red rectangle which is generated by a vision model (e.g., GPT-4o). The (M)LLM uses the query-specific annotation guidance in panel (c), along with the extracted and generated product attributes in panel (d), to assign a relevance label. In this example, as shown in panel (e), the label is ``highly\_relevant'', and the reasoning (aka the chain-of-thought step) of the (M)LLM is shown for inspection and debugging purposes.

\section{Human Annotation Guidelines}
\label{app:annotation_guidelines}

For human annotators, we focused on three classes:
\begin{itemize}
    \item \textbf{highly relevant}: The retrieved product satisfies all the specifications in the query.
    \item \textbf{acceptable substitute}: The item fulfils some, but not all aspects of the query and the retrieved item can be used as a functional substitute.
    \item \textbf{irrelevant}: A central aspect of the query is not fulfilled (e.g. wrong brand, wrong category, wrong product).
\end{itemize}
We decided against a more granular annotation scale to reduce mental load on annotators and to (hopefully) harness higher agreement scores among annotators. 

Our annotation guidelines also reflect requirements that are more business-specific rather than content-specific. For example, annotators have been explicitly briefed that if a query requests a specific brand (e.g. \emph{Polo Ralph Lauren jumpers}), any retrieved item that is not from the requested brand is to be regarded as ``irrelevant''. 

Another business-specific rule was that if a query requests a particular product (e.g. \emph{The North Face 1996 retro nuptse jacket}), any retrieved product that is not that particular type of The North Face jacket is to be regarded as ``irrelevant''.

Despite the explicit mentions of these rules, numerous provided examples across product categories, and an additional briefing session after the annotation pilot phase, brand and product errors were among the most commonly made human annotation errors.

\section{Experiments with LLM types: GPT-3.5, GPT-4, and GPT-4o}
\label{app:other-experiments-llm-type}

\begin{figure*}[ht]
\begin{center}
    \captionsetup{type=table}
    \caption{
    Agreements between (M)LLM and the human annotator groups (i.e., A1, A2 and tiebreaker). Similar to Table~\ref{tab:llm_vs_human_gpt4o_with_without}, but with additional columns showing the agreements of A1, A2, and the tiebreaker groups with other annotators. 
    We report agreements separately for English and German. For human annotations, we report the total time and cost. We use GPT-4o in all steps of our LLM annotation pipeline (Fig.~\ref{fig:pipeline}).
    }
    \includegraphics[width=1.0\linewidth]{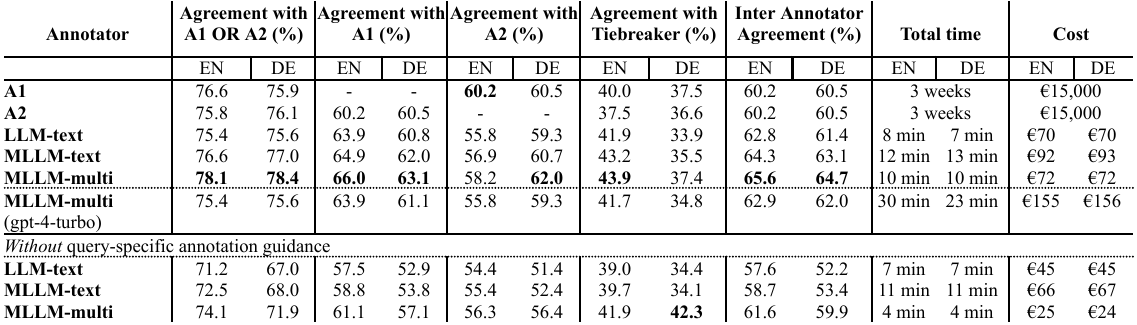}
    \label{tab:llm_vs_human_gpt4o_with_without_more_columns}
\end{center}
\end{figure*}

\begin{figure*}[ht]
\begin{center}
    \captionsetup{type=table}
    \caption{
    Like Table~\ref{tab:llm_vs_human_gpt4o_with_without_more_columns}, except we use GPT-3.5 Turbo (specifically, ``gpt-3.5-turbo-0125'') for text inputs and GPT-4o for generating textual descriptions for image inputs (Step~4 in Fig.~\ref{fig:pipeline}). Here, we do not have ``MLLM-multi'' as GPT-3.5 Turbo does not accept multimodal (text and image) inputs.
    }
    \includegraphics[width=1.0\linewidth]{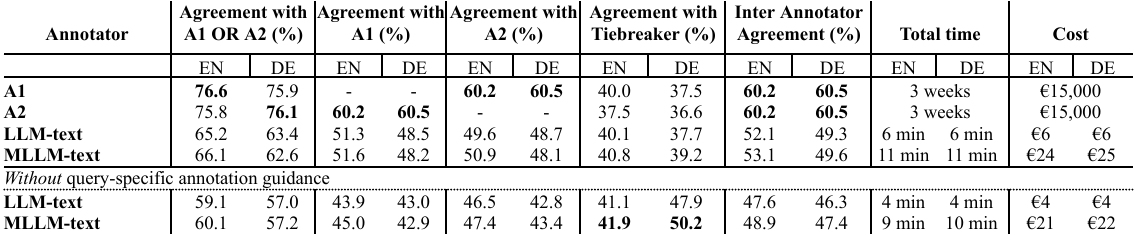}
    \label{tab:gpt35_4o_for_vision}
\end{center}
\end{figure*}

In this section, we compare human annotator groups with (M)LLMs using different architectures. The results in Table~\ref{tab:llm_vs_human_gpt4o_with_without_more_columns} are primarily based on ``GPT-4o'' \cite{openai_gpt4o}, except for the row labelled ``MLLM-multi (gpt-4-turbo)''. For GPT-4 Turbo, the agreement with human annotators was consistently lower than that of GPT-4o, while its costs and evaluation times exceeded those of all other architectures.\\
In Table~\ref{tab:gpt35_4o_for_vision}, we repeated the experiments using GPT-3.5~Turbo. As expected, its results were significantly worse compared to GPT-4o and GPT-4~Turbo, but its cost and time requirements were lower than those of the other architectures.

\section{LLM versus Human error types}
\label{app:llm_vs_human_error_types}

After manually inspecting a sample of hard disagreements\footnote{We consider a hard disagreement to be when, for example, the LLM considers a product to be ``highly relevant'' for a given query, whereas the human majority judgement would be ``irrelevant'', and vice versa. Refer to Section~\ref{sec:discussion} for details.}, we defined the following 9 error classes, some of which are applicable to LLMs and humans, and some to LLMs only:

\textbf{1. Brand error}. When a user specifies a brand name in the search query, e.g. \emph{Lee jeans}, \emph{Nike sneakers}, or \emph{Mascara dresses}, we consider any retrieved item as ``irrelevant'' if it is not from the requested brand. This is independent of whether the retrieved item would be visually similar to the requested one. This requirement has been covered in the LLM prompt as well as the human annotation guidelines. Predominantly, this error has been made by human annotators (see Fig.~\ref{fig:llm_vs_human_error_analysis}).

\textbf{2. Product error}. When a user specifies a specific product in the search query, e.g. \emph{Levis 501} or \emph{Adidas Stan Smith}, we consider any retrieved item that is not exactly the requested item as ``irrelevant''. This requirement has been covered in the LLM prompt as well as the human annotation guidelines. Predominantly, this error has been made by human annotators (see Fig.~\ref{fig:llm_vs_human_error_analysis}).

\textbf{3. Too strict}. This error happened when a product was judged as ``irrelevant'' for a given query despite fulfilling \emph{almost} all the requirements of the query. This error has been predominantly made by LLMs (see Fig.~\ref{fig:llm_vs_human_error_analysis}), for example when a query requested \emph{black Levis jeans with holes}, but the retrieved product was a grey pair of Levis jeans with holes, the LLM would typically annotate the retrieved products as ``irrelevant''.

\textbf{4. Too lenient}. This error happened when a product was judged as ``highly relevant'' for a given query, despite not fulfilling all requirements that the query specified. This error has been exclusively made by human annotators (see Fig.~\ref{fig:llm_vs_human_error_analysis}), for example where for a query like \emph{Nike Air Force One high-top}, humans annotated a \emph{Nike Air Force One low-top} sneaker as ``highly relevant''.

\textbf{5. Category error}. When a user specifies the category of a fashion item in the search query, e.g. \emph{dress}, \emph{sneakers}, \emph{belts}, we consider any retrieved item that does not match the category as ``irrelevant''. This requirement has been covered in the LLM prompt as well as the human annotation guidelines. Predominantly, this error has been made by human annotators (see Fig.~\ref{fig:llm_vs_human_error_analysis}).

\textbf{6. LLM hallucination error}. We rarely observed hallucinations as a source of error. Interestingly, when hallucinations did occur, they were exclusively related to size queries, such as \emph{t-shirt xxl}. In such cases, the LLM would hallucinate various available sizes for a given retrieved product and make a relevancy judgement on the basis of its hallucinations.

\textbf{7. LLM translation error}. Since our dataset contained German and English queries, the LLM was prompted to translate a German query into English before starting its reasoning process. This sometimes resulted in translation errors that subsequently led to incorrect relevancy judgements. For example, it happened for queries containing the term \emph{Unterziehhose}, meaning some sports leggings one can wear underneath sports shorts, which the LLM incorrectly translated as \emph{underpants}.

\textbf{8. LLM understanding error}. This error category is somewhat broader. We would categorise an LLM error as \emph{understanding error}, whenever the LLM misinterpreted a part of the query or the product. For example, this error occurred when the LLM would misinterpret a query for \emph{Nike Tech Fleece} to be focused on the material whereas \emph{Tech Fleece} typically refers to a particular series of Nike sports clothing. Another example is the misinterpretation of brand names, such as for \emph{On Vacation} (interpreted in its literal meaning), or for \emph{Evry Jewels} (where \emph{Evry} would be interpreted to mean \emph{Every}). To our amusement during error analysis, we also observed a brand misinterpretation for the  query \emph{miniature winter jackets for kids}, where \emph{Mini A Ture} is a kids' clothing brand. The LLM interpreted \emph{miniature} in its literal sense and reasoning that \emph{[...] the sizes available are for kids, which fits the 'miniature' requirement.}

\textbf{9. LLM vision error}. Some of our models included the visual interpretation of a product image in its relevancy assessment.\footnote{MLLM-text in Tables~\ref{tab:llm_vs_human_gpt4o_with_without},~\ref{tab:llm_vs_human_gpt4o_with_without_more_columns} and~\ref{tab:gpt35_4o_for_vision}.} We only rarely observed LLM vision errors. If they did occur, it was typically when the product image was taken at a slight angle—for instance, with a pair of sneakers where the LLM erroneously identified them as high-top due to the photo angle. Errors were also more likely when the image included a human model, which acted as a distractor.

\section{Subjective Nature of Relevance Judgements}
\label{app:subjectivity}
The difficulty in judging query-product relevancy can vary widely. For example, for queries such as \emph{Nike Air Max 95} or \emph{Paul Smith long sleeve polo shirt}, there is barely any room for subjectivity --- the retrieved products either are matches, or they are not. And indeed, this is reflected in the human-human inter-annotator agreement (95\% and 82\%, respectively) and the LLM-human agreement (98\% and 89\%, respectively), for these two examples.

However, there are numerous queries that are much more open to subjective judgement. One such example is the query \emph{smart casual shoes}, where the human-human agreement was only 12\% and the LLM-human agreement was 24\%. The range of suitable products for this query spans various types of shoes, and whether or not a particular shoe can be categorised as \emph{smart casual} is typically not included in the product data. In these cases, humans and LLMs would draw on their prior knowledge for making a relevance judgement. LLMs would generally be a stricter judge and consider anything that resembles a sneaker too closely, or is not in a shade of black or brown, as ``irrelevant''. Human strictness for relevancy judgements for this query varied between the very formal and the (loosely speaking) anything goes extremes.

\end{document}